\documentclass[preprintnumbers,showpacs,amsmath,amssymb,floatfix,prd,onecolumn,superscriptaddress,nofootinbib]{revtex4}

\usepackage{amsmath}
\usepackage{graphicx}
\usepackage{epsfig}
\usepackage{bm}
\usepackage{amsfonts}
\usepackage{epstopdf}
\usepackage{bm}
\usepackage[colorlinks,linkcolor=blue,backref]{hyperref}
\usepackage{cancel}

\begin{document}

\title{Photons generated by gravitional waves in the near-zone of a neutron star}

\author{Chao-Jun Feng}
\thanks{Corresponding author}
\email{fengcj@shnu.edu.cn}
\affiliation{Division of Mathematical and Theoretical Physics, Shanghai Normal University, 100 Guilin Road, Shanghai 200234,  P.R.China}

\author{Ao Guo}
\email{11912722@mail.sustech.edu.cn}
\affiliation{Department of Physics, Southern University of Science and Technology, Shenzhen 518055, P.R.China}

\author{Zhong-Ming Xie}
\email{11910518@mail.sustech.edu.cn}
\affiliation{Department of Physics, Southern University of Science and Technology, Shenzhen 518055, P.R.China}

\author{Miao Li}
\email{3498044240@qq.com}
\affiliation{Department of Physics, Southern University of Science and Technology, Shenzhen 518055, P.R.China}

\begin{abstract}
When a gravitational wave or a graviton travels through an electric or magnetic background, it could convert into a photon with some probability. In this paper, a dipole magnetic field 
is considered as this kind of background in both the Minkowski spacetime and the curved spacetime in the near-zone of a neutron star. In the former case, we find that the graviton traveling vertically rather than parallel to the background  magnetic field could be more effectively converted into an electromagnetic radiation field. In the latter case, we focus on the situation, in which the graviton travels along the radial direction near a neutron star. The radius of a neutron star is about ten kilometers, so the gravitational wave with long wavelength or low frequency may bypass neutron stars by diffraction. For  high frequency gravitational wave, the conversion probability is proportional to the distance square as that 
in the static electric or magnetic background case. The smaller the inclination angle between the dipole field and the neutron star north pole is, the larger  magnetic amplitude will be. The term that described  curved spacetime will slightly enhance this kind of probability. 
We estimate that this value is about the order of $\sim 10^{-14}- 10^{-10}$. Therefore, it is expectable that  this kind of conversion process may have a potential to open a window for observing high frequency gravitational waves.
\end{abstract}

%\pacs{14.80.bn, 98.80.Es, 98.80.Cq }

\maketitle
%\tableofcontents

\section{Introduction}

Field of high frequency gravitational wave(HFGW) is coming to vitality from both theoretical and experimental aspects. HFGW is theorized to be relic gravitational waves that are associated with imprints of the Big Bang like the CMB \cite{grishchuk1976primordial}. And at a very high frequency, the production of gravitational waves is possibly attributed to discrete sources, cosmological sources, brane-world Kaluza–Klein (KK) mode radiation, and plasma instabilities \cite{cruise2012potential}. Ground-based GW observations are sensitive to the low frequency such as in the range from 10Hz to kHz \cite{ejlli2019upper}. Novel ideas have been proposed to overcome difficulties and some of them are potential approaches to probing high or even ultra high frequency gravitational waves \cite{cruise2006prototype,akutsu2008search,cruise2012potential,ito2020probing,ejlli2019upper}.

When a gravitational wave propagates through a background electromagnetic field, it will slowly turn into a photon. The conversion of relic gravitational waves into photons in cosmological background magnetic fields is studied in \cite{dolgov2012conversion}, while the reverse process to probing how strong the primordial magnetic fields are generated is  discussed in  \cite{fujita2020gravitational}. In this paper, we will focus on the HFGW-photon conversion in the near-zone of a typical neutron star, which has a strong surface magnetic field in range of  $10^{8-15}$ G \cite{reisenegger2003origin} and mass of $1.2-2.0$ solar masses with radius of about 11km \cite{ozel_masses_2016}. A neutron star can be regarded as a rotating pulsar with a magnetic dipole field, and this field has an inclination angle with the north  pole of the neutron star\cite{kim_general_2021}. 

The gravitation wave or graviton coming into the neutron star can be from distance sources and propagate like a plane wave. Dark matter can be also one of these sources \cite{abbott2022all}. One can suppose that dark matter is distributed around a neutron star and gravitational waves are generated due to quantum fluctuations\cite{richard2015superradiance}. These GWs may convert to photons due to the neutron star's strong background magnetic field and then may be observed. In Section 2, the conversion process with plane gravitational waves coming from different directions under  a magnetic dipole background field in the Minkowski spacetime is considered. In Section 3, the electromagnetic field in the curved space-time background around the neutron star is reviewed, and then the conversion ratio with spherical gravitational waves propagating to a typical neutron star is calculated.  Section 4 is devoted to conclusions and discussions.

\section{Action and equations of motion}
When a gravitational wave coming from distance travels in the background of an electromagnetic field, it could slowly turns into a photon. The action describing such process is given by
\begin{eqnarray}\label{equ:action}
S = \int d^4x 
\sqrt{-\bar g}\Bigg[ -\frac{1}{4} f^{\mu\nu} f_{\mu\nu} + \bigg( \bar g^{\alpha\mu}\bar F^{\beta\nu} -\frac{1}{4}\bar g^{\alpha\beta}\bar  F^{\mu\nu}   \bigg)f_{\mu\nu} h_{\alpha\beta} \Bigg]\,,
\end{eqnarray}
where $\bar g_{\mu\nu}$ and $\bar F_{\mu\nu} $ are the background metric and the electromagnetic field, respectively. The electromagnetic field $f_{\mu\nu}$ is generated by the incoming gravitational wave (or the graviton) $h_{\mu\nu}$ when it is passing through the background $\bar F_{\mu\nu}$ field. Both $f_{\mu\nu}$ and $h_{\mu\nu}$ can be regarded as perturbations of the background fields as the following
\begin{eqnarray}
F_{\mu\nu} = \bar F_{\mu\nu} + f_{\mu\nu} \,,\quad g_{\mu\nu} = \bar g_{\mu\nu} + h_{\mu\nu}.
\end{eqnarray}
By definition, $f_{\mu\nu}$ is given by 
\begin{eqnarray}
	f_{\mu\nu} = \bar\nabla_\mu A_{\nu} - \bar\nabla_\nu A_{\mu} = \partial_\mu A_{\nu} - \partial_\nu A_{\mu}\,,
\end{eqnarray}
with the electromagnetic potential $A_{\mu}$. Here $\bar \nabla_\mu $ stands for the covariant derivative with the background metric $\bar g_{\mu\nu}$. The second term in the action (\ref{equ:action}) could be interpreted as an effective current $J_\mu$:
\begin{eqnarray}
S &=& \int d^4x 
\sqrt{-\bar g}\bigg[ -\frac{1}{4} f^{\mu\nu} f_{\mu\nu}+ \bar g^{\mu\nu} J_{\mu}A_{\nu}\bigg]\,.
\end{eqnarray}
That is, the equation of electromagnetic field satisfies:
\begin{equation}\label{equ:maxwell}
\bar\nabla_\mu f^{\mu\nu} = \frac{1}{\sqrt{-\bar g}}\partial_\mu (\sqrt{-\bar g}f^{\mu\nu}) = -J^\nu \,.
\end{equation}
After performing partial integration and omitting total derivatives, the effective current is identified as 
\begin{eqnarray} \label{equ:cur}
	J^\nu =  -\bar\nabla_\lambda  ( C^{\lambda\nu\alpha\beta}    h_{\alpha\beta}  ) \,,
\end{eqnarray}
where we have defined 
\begin{eqnarray} \label{equ:cq}
	C^{\lambda\nu\alpha\beta} \equiv \bar g^{\alpha\lambda} \bar F^{\beta\nu} -\bar g^{\alpha\nu}  \bar F^{\beta\lambda} -\frac{1}{2}\bar g^{\alpha\beta} \bar  F^{\lambda\nu} \,.
\end{eqnarray}
It is obvious that $C^{\lambda\nu\alpha\beta}$ is asymmetric  of the first two indexes $\lambda\nu$, i.e. $C^{\lambda\nu\alpha\beta} = - C^{\nu \lambda\alpha\beta} $.

\section{HFGW-photon conversion trough dipole magnetic field background in the Minkowski spacetime}

In this section, a gravitational wave is assumed to propagate in a flat spacetime, $\bar g_{\mu\nu} = \eta_{\mu\nu}$, then the effective current is:
\begin{eqnarray} \label{equ:curflat}
	J_\nu &=&-\bar{g}_{\alpha \nu} h_{\beta\mu} \partial^\mu  \bar F^{\beta\alpha} + \bar F^{\beta\mu}\partial_\mu h_{\nu\beta}   +h_{\nu\beta}\partial_\mu \bar F^{\beta\mu}  \,,
\end{eqnarray}
where the transverse-traceless (TT) gauge ($h=h^\mu_\mu=0,\partial_\mu h^\mu_\beta = 0$) is chosen. From the above equation, it is clearly that $J^0 = -h^\mu_\beta \partial_\mu \bar F^{\beta 0} $, which means charge can be only generated by an electric background field  along the $x$ or $y$ direction with non-zero gradient. The contribution from the graviton's variation is the second term $\sim\bar F^{\beta\mu}\partial_\mu h_{\nu\beta}$. In the TT gauge, there are two linearly independent polarization states, so the effective current density generated by the graviton has two components in a static field background.

\begin{figure}
	\centering
	\includegraphics[width=0.4\linewidth]{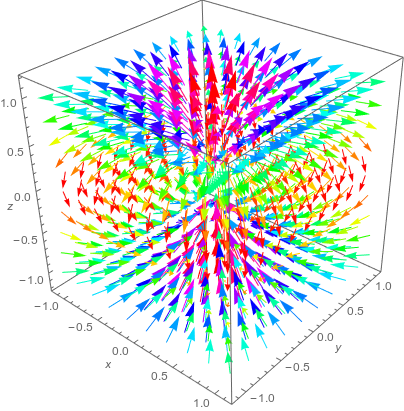}
	\caption{The vector diagram of a dipole magnetic field.}
	\label{fig:dipmag}
\end{figure}

Assuming that the $z$ axis is aligned with the dipole moment $\mathbf{m}$, then the background magnetic field is given by
\begin{eqnarray}
\mathbf{\bar B} = B_0\left( \frac{r_0}{r} \right)^3 \bigg( 2\cos\theta~\mathbf{e_r} + \sin\theta~\mathbf{e_\theta} \bigg)\,,
\end{eqnarray}
where $\mathbf{e_r}$ and $\mathbf{e_\theta}$ are unit coordinate vectors along the radial and the azimuthal directions in the spherical coordinates, respectively. Here $B_0$ is half of the magnitude in the polar direction of the dipole magnetic field, i.e. $B_0=B_p/2$ where $B_p  = |\mathbf{\bar B}(\theta = 0, r=r_0)$| . In the Cartesian coordinates this magnetic is
\begin{eqnarray}
\mathbf{\bar B} &=& B_0\left( \frac{r_0}{|\mathbf{x}|} \right)^3 \bigg( \frac{3zx}{|\mathbf{x}|^2}~\mathbf{e_x} +
\frac{3zy}{|\mathbf{x}|^2} ~\mathbf{e_y} +
\frac{2z^2-x^2-y^2}{|\mathbf{x}|^2}\ ~\mathbf{e_z} \bigg) \,, \label{equ:dipoleB}
\end{eqnarray}
and its modulus is
\begin{equation}
|\mathbf{\bar B}| = B_0\left( \frac{r_0}{|\mathbf{x}|} \right)^3 \sqrt{1+\frac{3z^2}{|\mathbf{x}|^2}}\,,
\end{equation}
with $\mathbf{e_x}, \mathbf{e_y}$ and $\mathbf{e_z}$ are the unit coordinate vectors and $|\mathbf{x}| = r$. To see the vector diagram of a dipole magnetic field, see Fig.\ref{fig:dipmag}.  From Equ.(\ref{equ:dipoleB}), it is easy to see that $\mathbf{\bar B}$ is symmetric about $x,y$ coordinates.

\subsection{A gravitational wave propagates along the $z$ direction}
\begin{figure}
	\centering
	\includegraphics[width=0.45\linewidth]{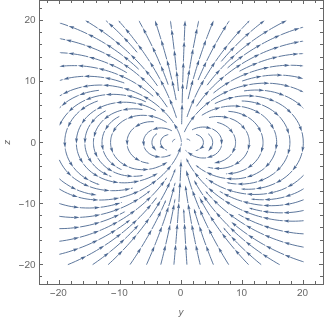}
	\caption{A dipole magnetic field projects on the $y-z$ plane.}
	\label{fig:yz}
\end{figure}

%\begin{figure}[h]
%	\centering
%	\includegraphics[width=0.4\linewidth]{dipolexys.png}
%	\includegraphics[width=0.4\linewidth]{dipolexyn.png}
%	\caption{The dipole magnetic field. Left: $z$ negative large, Right: $z$ positive large.}
%	\label{fig:yz}
%\end{figure}

%\subsubsection{In the case of  $h_{11}= - h_{22}$ mode}

At first, we consider that only one polarization mode (e.g. $h_{11}$) of the gravitational wave enters along the $z$ direction in the dipole field, see Fig.\ref{fig:yz}. This mode is described by
$h_{11} = -h_{22} = e_{11}e^{ik_\mu x^\mu} +  e_{11}^*e^{-ik_\mu x^\mu}$ with $k_\mu =(w,\mathbf{ k})= (w,k_x,k_y,k_z) =(w,0,0,k_z) $ and $w^2 = k_x^2+k_y^2+k_z^2 = |\mathbf{k}|^2$. Redefine the origin of time so that $e_{11}$ is real. In this case, the current density components are given by:
\begin{eqnarray}
J_x 
&=& -2|e_{11}| (iw\bar B_y + \partial_z \bar  B_y )e^{ik_\mu x^\mu} \,,\label{equ:zh11jx}\\
J_y
&=& -2|e_{11}| (iw\bar B_x + \partial_z \bar  B_x)e^{ik_\mu x^\mu}\,,\label{equ:zh11jy}\\
J_z&=&  4|e_{11}| \partial_x \bar  B_y e^{ik_\mu x^\mu}\,,\label{equ:zh11jz}
\end{eqnarray}
where the derivatives of $B_i$ can be calculated from  Equ.(\ref{equ:dipoleB}), see Sec.\ref{sec:appflat1} for details. 

In the far zone $|\mathbf{x}| \gg |\mathbf{x'}| $, the distance between the field and the source can be approximately by
\begin{eqnarray}
|\mathbf{x}-\mathbf{x}'| \approx |\mathbf{x}|-\frac{\mathbf{x}}{|\mathbf{x}|}\cdot \mathbf{x'} = |\mathbf{x}|-\mathbf{\hat{x}}\cdot \mathbf{x'}\,.
\end{eqnarray}
Then we have
\begin{eqnarray}\label{equ:approx}
-w(t-|\mathbf{x}-\mathbf{x}'|)+\mathbf{ k}\cdot \mathbf{x'} \approx 
 -w(t-|\mathbf{x}|)-w\mathbf{u}\cdot \mathbf{x'}\,,
\end{eqnarray}
with
\begin{equation}
\mathbf{u} \equiv \mathbf{\hat{x}}-\mathbf{ \hat k}\,, \quad \mathbf{ \hat k} \equiv \frac{\mathbf{ k}}{w}\,.
\end{equation}
The generated electric field  can be calculated by using the time derivative of the retarded potential: 
\begin{eqnarray}\label{equ:retar}
E_i(\mathbf{x},t) = -\partial_t A_i(\mathbf{x},t) = \frac{1}{4\pi }\int_{-L_x/2}^{L_x/2}\int_{-L_y/2}^{L_y/2}\int_{-L_z/2}^{L_z/2}  \frac{-\partial_t J_i(\mathbf{x}', t-|\mathbf{x}-\mathbf{x}'| )}{|\mathbf{x}-\mathbf{x}'| } dx'dy'dz'\,,
\end{eqnarray} 
where $L_x,L_y$ and $L_z$ indicate the range where the current is not zero. After straightforwardly calculation, the total power radiated can be estimated by
\begin{eqnarray}
\nonumber
P_{\gamma} = \frac{1}{2}\int |\mathbf{E}|^2 r^2 d \Omega \approx 
 \frac{3}{2}\pi |e_{11}|^2B_0^2 L^2\left(\frac{r_0}{L}\right)^6\left[ 1-\frac{1+4\xi^2}{(1+\xi^2)^4} \right]\,, 
\end{eqnarray}
with $\xi = R/L_z$ and $R=\sqrt{L_x^2+L_y^2}/2$, see Equ.(\ref{equ:pz11}) in Sec.\ref{sec:appflat2} for details. Here $L=2L_z$ can be regarded as the distance that the graviton travels along the $z$ direction.

The stress energy tensor of the graviton is 
\begin{eqnarray}
\langle t_{\mu\nu}\rangle = \frac{k_\mu k_\nu}{8\pi G}\left(|e_{11}|^2 + |e_{12}|^2\,, \right)\,.
\end{eqnarray}
which gives the power of the incoming gravtion as the following
\begin{eqnarray}
P_g = \frac{w^2|e_{11}|^2}{8\pi G} \int_0^R \rho d\rho \int_0^{2\pi} d\theta = \frac{w^2|e_{11}|^2}{8 G}  R^2 \,.
\end{eqnarray}

The probability for a graviton turning into photon in background electric field is estimated by the ratio of the power radiated to that of the incoming graviton:  
\begin{eqnarray} \label{equ:res1}
\epsilon_{g-\gamma} &=& \frac{P_{\gamma}}{P_g} = 12\pi G B_0^2L^2\left(\frac{r_0}{L}\right)^6\left(\frac{1}{wL}\right)^2\left[ \frac{\xi^2(6+4\xi^2+\xi^4)}{(1+\xi^2)^4}\right]\,,
\end{eqnarray}
which is much depressed because $wL\gg 1$ for a given source scale. For the $h_{12}= h_{21}$ mode , one can get the same  graviton energy transforming rate. 

\subsection{A gravitational wave propagates along the $y$ or $x$ direction}
Now we consider that a gravitational wave enters along the $y$ direction with $k_\mu=(w,0,k_y,0) $, i,e, $k_y = w$,  see Fig.\ref{fig:yz} for the background dipole magnetic field. The result will be the same for that along the $x$ direction because the magnetic filed is symmetric about $x,y$ coordinates.  
%\begin{figure}
%	\centering
%	\includegraphics[width=0.4\linewidth]{dipolexzfar.png}
%	\includegraphics[width=0.4\linewidth]{dipolexz.png}
%	\caption{The dipole magnetic field. Left: $y$ is large, Right: $y=0$.}
%	\label{fig:xz}
For the $h_{11}= - h_{33}$ mode, the current density components are:
\begin{eqnarray} 
J_x &=&2|e_{11}|( iw\bar B_{z} + \partial_y  \bar B_z)e^{ik_\mu x^\mu}\,,\label{equ:yh11jx}\\
J_y &=&  -4|e_{11}|\partial_x\bar B_ze^{ik_\mu x^\mu}\,,\label{equ:yh11jy}\\
J_z &=&2|e_{11}|(iw\bar B_x + \partial_y  \bar B_x  )e^{ik_\mu x^\mu} \,.\label{equ:yh11jz}
\end{eqnarray}
After straightforwardly calculation, the total power radiated can be estimated by
\begin{eqnarray}
\nonumber
P_{\gamma}&=& \frac{\pi}{2} w^2|e_{11}|^2B_0^2r_0^6L^2\left(\frac{4L^2+5 \epsilon ^2}{ \epsilon ^2 \left(L^2+\epsilon ^2\right)^2} - \frac{4L^2+5R ^2}{R ^2 \left(L^2+R ^2\right)^2}\right)
\approx 2\pi w^2|e_{11}|^2B_0^2r_0^4 \frac{r_0^2}{\epsilon^2} \,,
\end{eqnarray}
where $R = \sqrt{L_z^2+L_x^2}/2$. Here $\epsilon$ demotes the shortest radius that the graviton can travel in the $z-x$ plane, i.e. $x'^2+y'^2\geq \epsilon^2$ in the integration (\ref{equ:retar}) and $\epsilon < L, R$, see Equ.(\ref{equ:totpowery11}) in Sec.\ref{sec:appflat2} for details. Finally, the probability for a graviton turning into photon  is estimated by:  
\begin{eqnarray}
\epsilon_{g-\gamma} &=& \frac{P_{\gamma}}{P_g} = 16\pi G B_0^2r_0^2 \frac{r_0^4}{\epsilon^2 R^2}\,,
\end{eqnarray}
If the order of magnitude for $(\epsilon R)^{1/2}$ is the same as $r_0$, then 
\begin{eqnarray} \label{equ:res2}
\epsilon_{g-\gamma} \approx 4\pi G B_p^2r_0^2 \,.
\end{eqnarray}
For the $h_{13}= h_{31}$ mode , one can get the same  graviton energy transforming rate.

From the results (\ref{equ:res1}) and (\ref{equ:res2}) one can conclude that when 
a gravitational wave travels perpendicular to the background  magnetic field, it can be effectively converted into the electromagnetic radiation, otherwise when the propagation direction is parallel to the background  magnetic field, this effect is depressed,  see Fig.\ref{fig:yz} for the background magnetic field. This conclusion is consistent with that obtained from the static electromagnetic field. Therefore, in next section, we will focus on the perpendicular case, i.e. a gravitational wave is traveling along the radial direction of a neutron star.

\section{HFGW-photon conversion in the near-zone of a neutron star}
\subsection{Briefly review of  the magnetic field of a neutron star}

The spacetime outside a spherical rotating neutron star with a time-constant angular velocity can be described by the following metric:
\begin{equation}\label{equ:metriclow}
ds^2 = -N^2dt^2+\frac{1}{N^2}dr^2-2\Omega r^2\sin^2\theta dtd\phi +r^2d^2\theta + r^2\sin^2\theta d^2\phi \,,
\end{equation}
in the spherical coordinate system. Here $N^2 = 1-\frac{r_s}{r}$ and $r_s = 2M$ is the star's Schwarzchild radius. The non-diagonal component of the metric tensor leads  to dragging of the inertial frame of reference with an angular velocity $\Omega = 2J/r^3$ and here $J$ is the angular momentum. 

For an observer with a 4-velocity $u^\alpha$, the covariant components of the  electromagnetic field tensor are given by
\begin{eqnarray}
\bar F_{\alpha\beta} = 2u_{[\alpha}\bar E_{\beta] } + \epsilon_{\alpha\beta\gamma\delta}u^\gamma \bar B^\delta \,,
\end{eqnarray}
where $\bar E^\alpha$ and $\bar B^\alpha$ are the electric and magnetic four-vector fields, respectively. Here $T_{[\alpha\beta]} = (T_{\alpha\beta}-T_{\beta\alpha})/2$ and $\epsilon_{\alpha\beta\gamma\delta}$ is the pseudo-tensorial expression for the Levi-Civita symbol $\tilde\epsilon_{\alpha\beta\gamma\delta}$:
\begin{eqnarray}
\epsilon_{\alpha\beta\gamma\delta} = \sqrt{-g} \tilde\epsilon_{\alpha\beta\gamma\delta} \,,\quad \epsilon^{\alpha\beta\gamma\delta} = -\frac{1}{\sqrt{-g}} \tilde\epsilon_{\alpha\beta\gamma\delta}\,.
\end{eqnarray}
The magnetic field of neutron star is caused by the perfect fluid interior region of itself rather than its rotation, while the electric field is caused by both the magnetic field and the rotation of a neutron star. It can be found that  the strength of the electric field is much smaller than that of the magnetic field, i.e. $E\sim \Omega \times B$. Therefore, we will keep only linear terms for the angular velocity and ignore the background electric field in the following.

An observer is called a zero angular momentum observer (ZAMO) if he/she is  locally stationary (at fixed values of $r$ and $\theta$) but is dragged into rotation with respect to a reference that fixed with respect to a distant observer. The ZAMO has the following 4-velocity components
\begin{eqnarray}
u^\alpha = N^{-1}(-1,0,0,\Omega)\,,\quad u_\alpha = N(-1,0,0,0)\,,
\end{eqnarray}
up to the first order of $\Omega$. For ZAMOs, the electromagnetic nonzero components are
\begin{eqnarray}
\bar F_{10} &=& -\bar F_{01} = \Omega N^{-1}  r\sin\theta  \bar B^{2} \,,\\
\bar F_{20} &=& -\bar F_{02}= - \Omega  r^2\sin\theta  \bar B^{1} \,,\\
\bar F_{12} &=&  -\bar F_{21} =  N^{-1} r \bar B^{3} \,,\\
\bar F_{23} &=&  -\bar F_{32} = r^2\sin\theta \bar B^{1} \,,\\
\bar F_{13} &=& -\bar F_{31} =   -N^{-1} r\sin\theta \bar B^{2} \,,
\end{eqnarray}
and
\begin{eqnarray}
\bar F^{12} &=& -\bar F^{21} = \frac{N}{r} \bar B^{3}  \,,\\
\bar F^{13} &=& -\bar F^{31} = -\frac{N}{r\sin\theta} \bar B^{2} \label{equ:f31} \,,\\
\bar F^{23} &=& -\bar F^{32} =  \frac{1}{r^2\sin\theta}\bar B^{1}  \,.
\end{eqnarray}
where the magnetic field outside the neutron star is assumed to be a dipolar
field with components observed in the zero angular momentum frame as the following \cite{rezzolla_general_2001}
\begin{eqnarray}
\bar B^{1} &\equiv& \bar B^{\hat r} =  \hat f(r)\big(\cos\chi\cos\theta+\sin\chi\cos\phi\sin\theta\big) \,, \label{equ:B1}\\
\bar B^{2}&\equiv&\bar B^{\hat\theta} =  \hat g(r)\big(\cos\chi\sin\theta-\sin\chi\cos\phi\cos\theta\big) \,,\label{equ:B2}\\
\bar B^{3} &\equiv&\bar B^{\hat\phi}= \hat g(r)\sin\chi\sin\phi \,,\label{equ:B3}
\end{eqnarray}
where $\chi$ is the angle between the magnetic moment and the polar axis of a neutron star, and 
\begin{eqnarray}
\hat f(r) &=& -\frac{3}{8}B_0R^3 \frac{1}{M^3} \bigg[\ln N^2 +\frac{2M}{r}\left(1+\frac{M}{r}\right)\bigg]\,, \\
\hat g(r) &=&\frac{3}{8}B_0R^3\frac{N}{M^2r}\bigg( \frac{r}{M}\ln N^2+\frac{1}{N^2}+1\bigg)\label{equ:gr}\,,
\end{eqnarray}
are the relativistic corrections. Here $R$ is the radius of the neutron star and $B_0$  is the  value of the magnetic field in the polar direction in the Newtonian limit. Notice that the solution (\ref{equ:B1})-(\ref{equ:B3}) is  based on a hypothesis that there is no matter outside the star, i.e. $\bar \nabla_{\mu} \bar F^{\mu\nu} = 0$.

\subsection{Short wave approximation}
The explicit form for $\bar \nabla_\mu f^{\mu\nu}$ in  Equ.(\ref{equ:maxwell}) is given by
\begin{eqnarray}\label{equ:maxwell2}
\bar \nabla_\mu f^{\mu\nu} = \frac{1}{\sqrt{-\bar g}}\partial_\mu \bigg(\sqrt{-\bar g}\bar g^{\mu\alpha}\partial_{\alpha}A^\nu\bigg)
-\frac{1}{\sqrt{-\bar g}}\partial_\mu \bigg(\sqrt{-\bar g}\bar g^{\nu\alpha}\partial_{\alpha}A^\mu\bigg)\,.
\end{eqnarray}
Here the gravitational field outsider a neutron star can be regarded as two parts:
\begin{equation}
\bar g_{\mu\nu} = \eta_{\mu\nu} + \epsilon_{\mu\nu} \,,
\end{equation}
where $\epsilon_{\mu\nu} \sim ( 2M/R, J/M^2) $. Make a simple estimate, one can get $2M/R \approx 0.32\sim0.53$ for a typical neutron star, and $J/M^2\leq 0.1$ for the fastest-known millisecond pulsar PSR 1937+214\cite{rezzolla_general_2001}. The change of the background metric is assumed to be not faster than that of the generated electromagnetic radiation field. In other words, when the length scale $L$ over which the background varies (i.e.  $\partial_{\mu} \bar g_{\alpha\beta} /\bar g_{\alpha\beta} \sim \mathcal{O}(1/L)$) is much larger than the wavelength $\lambda_{EM}$ of the generated electromagnetic radiation field $L\gg \lambda_{EM}$, the term $\partial_{\mu}(\sqrt{-\bar g}\bar g^{\nu\alpha}\partial_\alpha A^\mu)$ in Equ.(\ref{equ:maxwell2}) can be approximated by $\bar g^{\nu\alpha}\partial_\alpha\partial_{\mu}(\sqrt{-\bar g} A^\mu)$, which will be vanished after taking the Lorentz gauge $\bar \nabla_{\mu} A^\mu  = 0$.  This is called the short wave approximation, which is a reasonable approximation when considering high-frequency wave. Then we can use the Green's function in curved spacetime to obtain the retarded potential field as the following \cite{dai_greens_2012}: 
\begin{eqnarray}
A^{\mu}(r,t) = \frac{1}{4\pi }\int  J^\mu(r', t') \psi_s(|r-r'|)dV'\,,
\end{eqnarray} \label{equ:ret}
where $\gamma$ is the determinant of the space metric, i.e. $\gamma = r^4\sin^2\theta /N^{2}$ and 
\begin{eqnarray}
\psi_s(r) &=& \int_r^{\infty} \sqrt{\frac{g_{rr}(R)}{-g_{tt}(R)}}\frac{dR}{R^2} = -\frac{\ln\left(1-\frac{2M}{r}\right)}{2M} \,,\\
t' &=& t-\int_{r'}^{r}\sqrt{\frac{g_{rr}(R)}{-g_{tt}{(R)}}}dR = t-(r-r')+2M\ln \frac{r-2M}{r'-2M} \,.
\end{eqnarray}
At large radius, the static function $\psi_s$ will reduce to the solution in a flat space $\psi_s\sim 1/r$, and $t'\sim t-(r-r')$, see Ref.\cite{dai_greens_2012} for details. Furthermore, the variation of the background electromagnetic field is  almost at the same order of that of the background metric, i.e. $\partial_\mu \bar F/\bar F\sim \mathcal{O}(1/L) $, then the effective current (\ref{equ:cur})  is approximated by 
\begin{eqnarray} \label{equ:curr}
J^\nu(r,\theta,\phi,t) &=& 
\nonumber
 -\bar\nabla_\lambda  ( C^{\lambda\nu\alpha\beta}    h_{\alpha\beta}  ) 
=  -\bar\nabla_\lambda  \bigg[ (  \bar g^{\alpha\lambda} \bar F^{\beta\nu} -\bar g^{\alpha\nu}  \bar F^{\beta\lambda} -\frac{1}{2}\bar g^{\alpha\beta} \bar  F^{\lambda\nu})h_{\alpha\beta}  \bigg]\\
&\approx& -\bar F^{\beta\nu}\bar g_{\alpha\beta} \partial_{\lambda}h^{\lambda\alpha}  + \bar F^{\beta\lambda}\bar g_{\alpha\beta} \partial_{\lambda} h^{\nu  \alpha} \,,
\end{eqnarray}
where we have used the transverse trace-less gauge condition. Note that the first term of the above equation can be eliminated under the harmonic gauge condition.

\subsection{A gravitational wave travels along the radial direction}

A gravitational wave travels along the radial  direction ($r$) with two polarization modes is described by 
\begin{eqnarray}
h^{22} = -\frac{\bar{g}_{33}}{\bar{g}_{22}} h^{33} =  e^{22}\frac{R}{r}e^{ik_\mu x^\mu} +  e^{22*}\frac{R}{r}e^{-ik_\mu x^\mu}\,,
\end{eqnarray}
and
\begin{eqnarray}
h^{23} = h^{32} =  e^{23}\frac{R}{r}e^{ik_\mu x^\mu} +  e^{23*}\frac{R}{r}e^{-ik_\mu x^\mu}\,,
\end{eqnarray}
with $x^\mu = (t,r,\theta,\phi)$, $k_\mu =(-w,\mathbf{ k})= (-w,w,0,0)  $ and $k^\mu = (w,w,0,0)  $. Here $R$ denotes some length scales. For example, this could be the radius length scale of a neutron star. Note that $e^{22}, e^{23}$ can be real by redefining the origin of the time. Then these modes can be expressed in a complex formalism
\begin{eqnarray}
h^{22}=2e^{22}\frac{R}{r}e^{ik_\mu x^\mu} \,,\quad h^{33} = - \frac{\bar{g}_{22}}{\bar{g}_{33}} h^{22}\,, \quad  
h^{23}=h^{32} = 2e^{23}\frac{R}{r}e^{ik_\mu x^\mu}\,,
\end{eqnarray}
which satisfies $\partial_\lambda h^{\lambda\beta}\approx 0$. The index values of non-zero items in the second term of Equ.(\ref{equ:curr}) are $\lambda=1,2$ because $\bar{F}^{\beta 0} = 0$ and $h_{\alpha \beta }$ doesn't depend on $\phi$. Then the effect current (\ref{equ:curr}) is given by  
\begin{eqnarray}
J^\nu(r,\theta,\phi,t) = K_{i}\bar F^{\beta i}\bar g_{\alpha\beta}  h^{\nu  \alpha}  \approx f^\nu(r,\theta,\phi) e^{-iw(t-r)}\,,
\end{eqnarray} 
where we have defined
\begin{eqnarray}\label{equ:fdef}
f^\nu(r,\theta,\phi) \equiv  2K_1\bar F^{\beta 1}\bar g_{\alpha\beta} \frac{R}{r} e^{\nu \alpha }\,.
\end{eqnarray}
Here we have used
\begin{eqnarray}
\partial_{\lambda}h^{\alpha  \beta} = \left(i k_\lambda  -\frac{1}{r}\delta_{\lambda,1}\right)h^{\alpha\beta} \equiv K_\lambda h^{\alpha\beta} \,,
\end{eqnarray}
with
\begin{eqnarray}
K_0 = -iw \,,\quad K_1 = iw-\frac{1}{r} =-K_0\left(1-\frac{1}{iwr}\right) \,,\quad  K_2 \approx 0,\quad  K_3=0\,.
\end{eqnarray}
Then the retarded potential is
\begin{eqnarray}
	A^{\mu}(\mathbf{r},t) &=& \frac{1}{4\pi }\int  J^\mu(\mathbf{r'}, t') \psi_s(|\mathbf{r}-\mathbf{r'}|)dV'
	\approx \frac{e^{-iw(t-r)}}{4\pi r } I^\mu  \,,\label{equ:poten}
\end{eqnarray} 
where $\mathbf{r}=(r,\theta,\phi)$ , $r=|\mathbf{r}|$,
\begin{eqnarray}
t'\approx t-|\mathbf{r}-\mathbf{r'}| \approx t-r+ \mathbf{\hat r}\cdot  \mathbf{r'}\,,\quad \mathbf{\hat r} = \frac{\mathbf{r}}{r}\,,
\end{eqnarray}
 and 
\begin{eqnarray}
I^\mu &=&  \int  f^\mu(\mathbf{r'}) e^{-iw(\mathbf{\hat r}\cdot \mathbf{\hat r'}-1)r'} dV'=\int  \sqrt{\gamma'} f^\mu(\mathbf{r'}) e^{-iw(\mathbf{\hat r}\cdot \mathbf{\hat r'}-1)r'} dr'd\theta'd\phi'\,,\label{equ:int2}
\end{eqnarray}
with   $\gamma'= r'^4\sin^2\theta' N^{-2}$ the determinant of the space metric.  From Equ.(\ref{equ:fdef}), we get
$f^0=f^1 =0$, which leads to $A^0 = A^1 = 0$ by using Equ.(\ref{equ:poten}).

The  components of the generated electric and magnetic radiation field are calculated by
\begin{eqnarray}
E^i(\mathbf{r},t)= F^{0i}\,,\quad 
B_i(\mathbf{r},t) = \frac{1}{2}\epsilon_{ijk} F^{jk}\,,
\end{eqnarray}
which gives $E^1=B^1= 0$ and 
\begin{eqnarray}
E^2 &=& \partial^0A^2(\mathbf{r},t) = K_0\bar g^{00} A^2\,, \\
E^3 &=& \partial^0A^3(\mathbf{r},t) = K_0\bar g^{00} A^3 \,,\\
B^2 &=& \partial^1A^3(\mathbf{r},t) = -K_1 \sqrt{\gamma} \bar g^{11}\bar g^{22}A^3\,,\\
B^3 &=& \partial^1A^2(\mathbf{r},t) = K_1\sqrt{\gamma}\bar g^{11}\bar g^{33}A^2\,,
\end{eqnarray}
where $\epsilon_{ijk}=\sqrt{\gamma}\tilde{\epsilon}_{ijk}$. Here we have used the relations $\partial_\nu A^\mu = K_\nu A^\mu$ from Equ.(\ref{equ:poten}). Then we get the Poynting vector $\mathcal{S}_i = \text{Re}( \frac{1}{4} \epsilon_{ijk} E^{j*} B^{k})$ which has only one nonzero component:
\begin{eqnarray*}
\mathcal{S}_1 &=& \frac{\sqrt{\gamma}}{2}\text{Re}(E^{2*}B^{3}-E^{3*}B^{2})\\
&=& \frac{1}{2}\text{Re}\bigg[ K_0^*K_1\gamma \bar g^{00}\bar g^{11}\bigg(  \bar g^{33}  |A^2|^2 + \bar g^{22}  |A^3|^2 \bigg) \bigg] \\
&=&
-\text{Re}\bigg[ \frac{K_0^*K_1}{32\pi r^2}\gamma\bigg(  \bar g^{33}  |I^2|^2 + \bar g^{22}  |I^3|^2 \bigg) \bigg]\,. 
\end{eqnarray*}

The total power radiated is estimated by 
\begin{eqnarray}
P_\gamma &=& \int |S_1| r^2d\Omega = \frac{ 1}{32\pi^2} \int  \gamma |K_0^*K_1|\bigg( \bar g^{33}  |I^2|^2 + \bar g^{22}  |I^3|^2 \bigg)d\Omega \,,\label{equ:int}
\end{eqnarray}
where we only keep the terms up to the first order of $1/r$ in $S^1$ because terms with high orders of $1/r$ in the electric field can not travel to distant observers.

Note that in Equ.(\ref{equ:int2}) there is an integral of $\phi$ in range of $[0,2\pi]$, $\theta$ in range of $[0,\pi]$, so the terms that proportional to $\sin\phi$, $\cos\phi$ and $\cos\theta$ in Equ.(\ref{equ:fdef}) will be vanished after performing the integration. Therefore, these terms can be dropped safely to get
\begin{eqnarray}
f^2(r,\theta,\phi) &=& 2K_1\bar F^{3 1}\bar g_{33} \frac{R}{r} e^{23 } =
2K_1\bar F^{3 1}\bar g^{22} \frac{R}{r} e_{23}
  \sim G(r) e_{23}\,,\\
f^3(r,\theta,\phi) &=& 2K_1\bar F^{3 1}\bar g_{33} \frac{R}{r} e^{33} =-
2K_1\bar F^{3 1}\bar g_{22} \frac{R}{r} e^{22} \sim - G(r) e_{22}\,,
\end{eqnarray}
where  we have defined
\begin{eqnarray}
G(r) \equiv   %K_1\frac{R}{r}\bar F^{3 1}\bar g^{22} 
%= 
\frac{3}{4}K_1  B_0R^4\cos\chi  \frac{N^2}{M^2r^5}\bigg( \frac{r}{M}\ln N^2+\frac{1}{N^2}+1\bigg) \,,
\end{eqnarray}
and used Equs.(\ref{equ:f31}), (\ref{equ:B2}) and (\ref{equ:gr}). 
Then the power can be expressed by
\begin{eqnarray}
P_\gamma \approx \frac{1}{32\pi^2}  \bigg(|e_{23}|^2I^{G1} + |e_{22}|^2I^{G2}\bigg) \,,
\end{eqnarray}
where we have also defined
\begin{eqnarray}
I^{G1}=\int \bar{g}^{33}(r')\gamma' \sqrt{\gamma'}K_1 G^*(r')dr'd\theta'd\phi'  \int \sqrt{\gamma''}  K_0^*G(r'') I^{\Omega}
 dr''d\theta''d\phi'' \,, \\
I^{G2}=\int \bar{g}^{22}(r')\gamma' \sqrt{\gamma'} K_1G^*(r') dr'd\theta'd\phi' \int \sqrt{\gamma''}  K_0^*G(r'') I^{\Omega}
 dr''d\theta''d\phi'' \,,
\end{eqnarray}
and
\begin{eqnarray}\label{equ:iomeg}
I^\Omega = \int e^{-iw(\mathbf{\hat r}\cdot \mathbf{\hat r'}-1)r' + iw(\mathbf{\hat r}\cdot \mathbf{\hat r''}-1)r''} \sin\theta d\theta d\phi\,.
\end{eqnarray}
By using  the short wave approximation ($wL\gg1$), we obtain
\begin{eqnarray}
	\label{equ:iomeg1}
I^{G1}\approx\frac{4\pi^2}{w^2} \int  r'^4N^{-2} K_1G^*(r') r''^2 dr'\int  N^{-1} K_0^*G(r'') 
dr'' \sin\theta' d\theta'd\phi' \,,\\
\label{equ:iomeg2}
I^{G2}\approx\frac{4\pi^2}{w^2} \int  r'^4N^{-2} K_1G^*(r') r''^2  dr'\int N^{-1} K_0^*G(r'') 
dr'' \sin^3\theta' d\theta'd\phi' \,.
\end{eqnarray}
These two integrals can be  calculated straightforwardly, see Sec.\ref{sec:app2} for details, and they are give by  
\begin{eqnarray}
I^{G1} &=&  16\pi w^2\pi^2  \cos^2\chi \frac{B_0^2L^2}{R^2} \left(1 + \frac{3M}{R}\right)^2 \,,\\
I^{G2} &=&  \frac{32\pi }{3}w^2\pi^2  \cos^2\chi \frac{B_0^2L^2}{R^2} \left(1 + \frac{3M}{R}\right)^2\,,
\end{eqnarray}
so the power is 
\begin{eqnarray}
P_\gamma \approx \frac{\pi w^2 }{2}  \bigg(|e_{23}|^2 +\frac{2}{3} |e_{22}|^2\bigg)  \cos^2\chi \frac{B_0^2L^2}{R^2} \left(1 + \frac{3M}{R}\right)^2  \,,
\end{eqnarray}
up to the leading order of $M/R$. Here $L$ denotes the distance that a gravitational wave travels along the radial direction, and we have also taken the approximation $L\ll R$ in the above results.

The averaged energy momentum tensor of the gravitational wave in a curved spacetime can be estimated by
\begin{eqnarray}
t_{\mu\nu} = \frac{1}{32\pi G} \partial_\mu h_{\alpha\beta}\partial_\nu h^{\alpha\beta }\,,
\end{eqnarray}
in which the energy flux along the $r$ direction is $t_{10}$. Then the power of the traveling gravitational wave is 
\begin{eqnarray}
	\nonumber
P_g &=& \int \frac{|K_1K_0^*|}{32\pi G}\bigg((\bar{g}^{22})^2|e_{22}|^2 + \bar{g}^{22}\bar{g}^{23}|e_{23}|^2\bigg) \frac{R^2}{r^2} r^2\sin^2\theta  d\Omega \\
%&=&\int  \frac{|K_1K_0^*|}{8\pi G}\bigg(\sin^2\theta |e_{22}|^2 + |e_{23}|^2\bigg)   \frac{R^2}{r^4} \sin\theta  d\theta d\phi  \\
%&=&\frac{|K_1K_0^*|}{8\pi G}  \frac{R^2}{r^4}\bigg(\frac{8\pi}{3} |e_{22}|^2 + 4\pi|e_{23}|^2\bigg)  \\
%&=&
%\frac{|K_1K_0^*| }{2G}  \frac{R^2}{r^4}\bigg( \frac{2}{3}|e_{22}|^2 +  |e_{23}|^2\bigg) \\
&\approx& \frac{ w^2}{2GR^2}  \bigg(\frac{2}{3}|e_{22}|^2 + |e_{23}|^2\bigg)\,. 
\end{eqnarray}
Finally, the probability for a graviton turning into a photon in the near zone of a neutron star is estimated by the ratio of the power radiated to that of the traveling graviton:  
\begin{eqnarray}\label{equ:resmain}
\epsilon_{g-\gamma} &=& \frac{P_{\gamma}}{P_g} \approx  \pi  GB_0^2L^2  \cos^2\chi  \left(1 + \frac{3M}{R}\right)^2  \,,
\end{eqnarray}
in the leading order of $M/R$ and $L\ll R$. 
From Equ.(\ref{equ:resmain}), the probability is proportional to the distance square, which is the same as that in the static electric or magnetic background case. The smaller of the inclination angle $\chi$, the larger of the magnetic amplitude, then this probability will be larger. The term caused by the curved spacetime is proportional to $(1+3M/R)^2$, which  will slightly enhance this kind of probability.  

The measured surface magnetic field of $\sim 1.6\times 10^{13}$G for Swift J0243.6+6124 is the strongest for all known neutron starts with detected electron cyclotron resonance scattering features\cite{Kong:2022cbk}.
By taking the values of $M/R\sim 0.21$, $\chi=60^\circ$ from \cite{Zhuravlev:2021fvm} and $L\sim 1$km, one can estimate the probability as follows
\begin{eqnarray}
\epsilon_{g-\gamma} \approx  2.1\times 10^{-38}\left(\frac{B_0L}{\text{T}\cdot \text{m}}\right)^2 \cos^2\chi\left(1+\frac{3M}{R}\right)^2 
%\approx 2.1\times 10^{-38} (1.6\times 10^{9}\cdot 10^3)^2 \cdot (1+3*0.21)^2 
\approx 3.57\times 10^{-14}\,.
\end{eqnarray}
For a magnetar with a higher magnetic field, such as $\sim 10^{15}$G, the order of magnitude of this conversion rate will reach $\sim 10^{-10}$. Compare another situation, in which gravitons travel in the universe between galaxies. There are more gravitons in the universe than we
believed. A typical galaxy has a size of $100000$ light years ($10^{21}$m) and an average magnetic field of $10^{-9}$ T, and
most of the fields are not turbulent, the conversion probability is on the order of $10^{-14}$ (versus $\sim 10^{40}$ for thermal radiation). Therefore, it has a greater opportunity to detect HFGWs through this conversion phenomenon in the near-zone of a neutron star or even a magnetar.

\section{Conclusion}
In this paper, we calculate the ratio of graviton-photon conversion under the backgrounds of a dipole magnetic field in a Minkowski spacetime and a typical neutron star magnetic field in a slightly rotational curved spacetime. In the former case, we find that when a GW travels perpendicular to the background  magnetic field, it can be effectively converted into an electromagnetic radiation field, but this conversion will be depressed if the GW travels  parallel to the background  magnetic field; so in the latter case, we focus on the case of that a GW travels along the radial direction near a neutron star. The conversion probability is proportional to the distance square, which is the same as that in the static electric or magnetic background case. The smaller the inclination angle $\chi$ is and the larger the magnetic amplitude is, the higher this probability will be. The term caused by the curved spacetime is proportional to $(1+3M/R)^2$, which  will slightly enhance this kind of probability. This conversion probability is on the order of $\sim 10^{-14}- 10^{-10}$ for a neutron star or a magnetar.

In recent years, some new ideas for searching for dark matters through GWs have been raised\cite{vermeulen2021direct}. A graviton that could convert into a photon in the near-zone of a typical neutron star will have a frequency  $ \nu \gtrsim c/R\sim 2.7\times 10^7$Hz. Therefore, this conversion process may open a window to observe HFGW, and we believe that to further study on a high-frequency GW and its conversion process is worthwhile. On the other hand, with the assistance of gravitational waves, it is also possible to enlighten a new effective theorem and a new observation method for dark matters.

\acknowledgments
This work is supported by National Science Foundation of China grant Nos.~12175099 and~11105091.

\appendix
\section{Detail calculations in the Minkowski spacetime}

\subsection{ The background dipole magnetic field and the effective current} \label{sec:appflat1}
From Equ.(\ref{equ:dipoleB}), the derivatives of background magnetic field with respect to coordinates are given by 
\begin{eqnarray}
\partial_{z}\bar B_x &=& \partial_{x}\bar B_z = 3B_0 r_0^3 \frac{x}{|\mathbf{x}|^5}\left(1-5\frac{z^2}{|\mathbf{x}|^2}\right) \,,\\
\partial_{z}\bar B_y &=& \partial_{y}\bar B_z =3B_0 r_0^3 \frac{y}{|\mathbf{x}|^5}\left(1-5\frac{z^2}{|\mathbf{x}|^2}\right)\,, \\
\partial_{x}\bar B_y &=& \partial_{y}\bar B_x = -15B_0 r_0^3\frac{xyz}{|\mathbf{x}|^7} \,,\\
\partial_{x}\bar{B}_x &=& 3B_0 r_0^3 \frac{z}{|\mathbf{x}|^5}\left(1-5\frac{x^2}{|\mathbf{x}|^2}\right) \,,\\
\partial_{y}\bar{B}_y &=& 3B_0 r_0^3 \frac{z}{|\mathbf{x}|^5}\left(1-5\frac{y^2}{|\mathbf{x}|^2}\right) \,,\\
\partial_{z}\bar{B}_z &=& -3B_0 r_0^3 \frac{z}{|\mathbf{x}|^5}\left(1+5\frac{z^2}{|\mathbf{x}|^2}\right) \,.
\end{eqnarray}

%The current components can be obtained from Equ.(\ref{equ:curflat}) as 
%\begin{eqnarray}
%\label{equ:zh11j1}
%J_x &=&\bar F^{12}\partial_2 h_{11}+\bar F^{13}\partial_3 h_{11}   +h_{11}\partial_3 \bar F^{13} \\
%\label{equ:zh11j2}
%J_y &=&\bar F^{12}\partial_1 h_{11}-\bar F^{23}\partial_3 h_{11}   -h_{11}\partial_3 \bar F^{23} \\
%\label{equ:zh11j3}
%J_z &=&- h_{11} \partial_1  \bar F^{13} + h_{11} \partial_2  \bar F^{23} \,.
%\end{eqnarray}
By using  the definition of $\bar F^{\mu\nu}$:
\begin{eqnarray}
\bar F^{12} &=& \bar B_z \,,\quad 
\bar F^{13} =  -\bar B_y \,, \quad 
\bar F^{23} = \bar B_x \,,
\end{eqnarray}
the current components can be obtained from Equ.(\ref{equ:curflat}) for the $h_{11}=-h_{22}$ mode  
%\begin{eqnarray}
%\label{equ:zh11j11}
%J_x &=&\bar B_z\partial_y h_{11}-\bar B_y\partial_z h_{11}   -h_{11}\partial_z \bar B_y\,, \\
%\label{equ:zh11j21}
%J_y &=&\bar B_z\partial_x h_{11}-\bar B_x\partial_z h_{11}   -h_{11}\partial_z \bar B_x \,, \\
%\label{equ:zh11j31}
%J_z &=& h_{11} \partial_x  \bar B_y + h_{11} \partial_y  \bar B_x \,.
%\end{eqnarray}
\begin{eqnarray}
\label{equ:zh11jx1}
J_x 
&=& 2|e_{11}| \bigg[ i(k_y\bar B_z-k_z\bar B_y) - \partial_z \bar  B_y \bigg]e^{ik_\mu x^\mu} \,,\\
\label{equ:zh11jy1}
J_y
&=& 2|e_{11}|\bigg[i(k_x\bar B_z-k_z\bar B_x) - \partial_z \bar  B_x \bigg]e^{ik_\mu x^\mu}\,,\\
\label{equ:zh11jz1}
J_z&=&   2|e_{11}| \bigg( \partial_x \bar  B_y +\partial_y \bar  B_x \bigg)  e^{ik_\mu x^\mu} = 4|e_{11}| \partial_x \bar  B_y e^{ik_\mu x^\mu}\,,
\end{eqnarray}
where we have used  $h_{11} = 2|e_{11}|e^{ik_\mu x^\mu}$ in the complex formalism and $\partial_i  h_{11} =ik_ih_{11} $, for $i=x,y,z$.

%\subsection{For the $h_{12}=h_{21}$ mode }\label{sec:appzh12}
%From Equ.(\ref{equ:curflat}), it is easy to get 
%\begin{eqnarray} 
%J_1 &=& -\bar F^{12}\partial_1 h_{12} +\bar F^{23}\partial_3 h_{12}   +h_{12}\partial_3 \bar F^{23}  \,,\\
%J_2 &=&   \bar F^{12}\partial_2 h_{12} + \bar F^{13}\partial_3 h_{12}  +h_{12}\partial_3 \bar F^{13}  \,,\\
%J_3 &=&- h_{12} \partial_2  \bar F^{13}- h_{12} \partial_1  \bar F^{23}  \,,
%\end{eqnarray}
%or
%\begin{eqnarray} 
%J_x &=& -\bar B_z\partial_x h_{12} +\bar B_x\partial_z h_{12}   +h_{12}\partial_z \bar B_x  \,,\\
%J_y &=&   \bar B_z\partial_y h_{12} - \bar B_y\partial_z h_{12}  -h_{12}\partial_z \bar B_y  \,,\\
%J_z &=&h_{12} \partial_y  \bar B_y- h_{12} \partial_x  \bar B_x  \,,
%\end{eqnarray}

The current components for the $h_{12}=h_{21}$ mode are 
\begin{eqnarray} 
	\label{equ:zh12jx1}
J_x &=&2|e_{11}|\bigg[ i(k_z\bar B_x-k_x\bar B_z )  +\partial_z \bar B_x  \bigg]e^{ik_\mu x^\mu} \,,\\
	\label{equ:zh12jy1}
J_y &=&2|e_{11}|  \bigg[  i(k_y\bar B_z-k_z\bar B_y )-\partial_z \bar B_y \bigg]e^{ik_\mu x^\mu}  \,,\\
	\label{equ:zh12jz1}
J_z &=&2|e_{11}|(  \partial_y  \bar B_y-  \partial_x  \bar B_x )e^{ik_\mu x^\mu} \,,
\end{eqnarray}
where we have used $h_{12} = 2|e_{11}|e^{ik_\mu x^\mu}$ and  $\partial_i  h_{12} = ik_ih_{12}$, for $i=x,y,z$.

%\subsubsection{with the $h_{11}=-h_{33}$ mode }\label{sec:appyh11}
%From Equ.(\ref{equ:curflat}), it is easy to get 
%\begin{eqnarray} 
%J_1 &=&\bar F^{12}\partial_2 h_{11}+\bar F^{13}\partial_3 h_{11}+
%h_{11} \partial_2  \bar F^{12} \,,\\
%J_2 &=& -h_{11}\partial_1\bar{F}^{12}-h_{11}\partial_3\bar{F}^{23} \,,\\
%J_3 &=&\bar F^{13}\partial_1 h_{11}+\bar F^{23}\partial_2 h_{11}+
%h_{11} \partial_2  \bar F^{23} \,,
%\end{eqnarray}
%or
%\begin{eqnarray} 
%J_x &=&\bar B_{z}\partial_y h_{11}-\bar B_y\partial_z h_{11}+
%h_{11} \partial_y  \bar B_z \,,\\
%J_y &=& -h_{11}\partial_x\bar B_z-h_{11}\partial_z\bar B_x \,,\\
%J_z &=&-\bar B_y\partial_x h_{11}+\bar B_x\partial_y h_{11}+
%h_{11} \partial_y  \bar B_x \,,
%\end{eqnarray}

The current components for the  $h_{11}=-h_{33}$ mode are 
\begin{eqnarray} 
		\label{equ:zh11jx2}
J_x &=&2|e_{11}|\bigg[ i(k_y\bar B_{z}-k_z\bar B_y) + \partial_y  \bar B_z\bigg]e^{ik_\mu x^\mu}\,,\\
\label{equ:zh11jy2}
J_y &=& -2|e_{11}|(\partial_x\bar B_z+\partial_z\bar B_x )e^{ik_\mu x^\mu} = -4|e_{11}|\partial_x\bar B_ze^{ik_\mu x^\mu}\,,\\
\label{equ:zh11jz2}
J_z &=&2|e_{11}|\bigg[i(k_y\bar B_x -k_x\bar B_y)+ \partial_y  \bar B_x  \bigg]e^{ik_\mu x^\mu} \,,
\end{eqnarray}
where we have used  $h_{11} = 2|e_{11}|e^{ik_\mu x^\mu}$ and $\partial_i  h_{11} = ik_i h_{11}$, for $i=x,y,z$.

%
%\subsubsection{with the $h_{13}=h_{31}$ mode}\label{sec:appyh13}
%From Equ.(\ref{equ:curflat}), it is easy to get 
%\begin{eqnarray} 
%J_1 &=& -\bar F^{13}\partial_1 h_{13}-\bar F^{23}\partial_2 h_{13}   -h_{13}\partial_2 \bar F^{23}  \,,\\
%J_2 &=&-h_{13} \partial_3  \bar F^{12} +h_{13} \partial_1  \bar F^{23} \,,\\
%J_3 &=& \bar F^{12}\partial_2 h_{31} +\bar F^{13}\partial_3 h_{31}  +h_{31}\partial_2 \bar F^{12}  \,,
%\end{eqnarray}
%or
%\begin{eqnarray} 
%J_x &=& \bar B_y\partial_x h_{13}-\bar B_x\partial_y h_{13}   -h_{13}\partial_y \bar B_x  \,,\\
%J_y &=&-h_{13} \partial_z  \bar B_z +h_{13} \partial_x  \bar B_x \,,\\
%J_z &=& \bar B_z\partial_y h_{31} -\bar B_y\partial_z h_{31}  +h_{31}\partial_y \bar B_z  \,,
%\end{eqnarray}
The current components for the  $h_{13}=h_{31}$ mode are 
\begin{eqnarray} 
	\label{equ:zh13jx2}
J_x &=&2|e_{11}| \bigg[  i(k_x\bar B_y-k_y\bar B_x ) -\partial_y \bar B_x  \bigg]e^{ik_\mu x^\mu} \,,\\
\label{equ:zh13jy2}
J_y &=&2|e_{11}|(\partial_x  \bar B_x- \partial_z  \bar B_z)e^{ik_\mu x^\mu} \,,\\
\label{equ:zh13jz2}
J_z &=& 2|e_{11}| \bigg[ i(k_y\bar B_z-k_z\bar B_y  ) +\partial_y \bar B_z \bigg]e^{ik_\mu x^\mu} \,,
\end{eqnarray}
where we have used  $h_{13} = 2|e_{11}|e^{ik_\mu x^\mu}$ and  $\partial_i  h_{13} = ik_i h_{13}$, for $i=x,y,z$.

\subsection{The generated electric field $E_i$  and its total power}\label{sec:appflat2}

The component of current density can be expressed as
\begin{eqnarray}
J_i(t,x,y,z) = f_i(x,y,z) e^{i k_\mu x^\mu } =  f_i(x,y,z) e^{i(\mathbf{k}\cdot\mathbf{x}-wt ) }\,,
\end{eqnarray}
with some function $f_i(x,y,z)$. By using Eqs.(\ref{equ:approx}) and (\ref{equ:retar}) , we get
\begin{eqnarray}
E_i(\mathbf{x},t)&=& -\frac{1}{4\pi |\mathbf{x}|}\int \partial_t J_i(\mathbf{x}', t-|\mathbf{x}-\mathbf{x}'| ) dV'\,,\\
&=& \frac{iwe^{ -iw(t-|\mathbf{x}|)} }{4\pi |\mathbf{x}|}\int  f_i(x',y',z') e^{ -iw\mathbf{u}\cdot \mathbf{x'} }  dV'\,,
\end{eqnarray}
where $dV'=dx'dy'dz'$. 
The total power radiated can be calculated by:
\begin{eqnarray}
\nonumber
P &=& \frac{1}{2}\int |\mathbf{E}|^2r^2d\Omega = \frac{1}{2}\sum_{i=1}^{3}\int |E_i|^2r^2d\Omega \\
\nonumber
&=& \frac{w^2 }{32\pi^2} \sum_{i=1}^{3}\int d\Omega \int  f_i(x',y',z') e^{ -iw\mathbf{u}\cdot \mathbf{x'} }  dV' \int  f_i^*(x'',y'',z'') e^{ iw\mathbf{u}\cdot \mathbf{x''} } dV''\\
&=&  \frac{w^2 }{32\pi^2}\sum_{i=1}^{3} \int d\Omega   f_i(x',y',z')f_i^*(x'',y'',z'') e^{ -iw(\mathbf{\hat{x}-\hat k})\cdot (\mathbf{x'}-\mathbf{x''}) }  dV'   dV'' \,. \label{equ:totpower}
\end{eqnarray}

\subsubsection{A gravitational wave propagates along the $z$ direction}\label{sec:gwz}

In this case, $k_z = w$ and $\hat k_z = 1$, then the total power radiated can be estimated from Equ.(\ref{equ:totpower}) as the following
\begin{eqnarray}
P&=& \frac{w^2 }{32\pi^2} \sum_{i=1}^{3}\int   f_i(x',y',z')f_i^*(x'',y'',z'') I(\mathbf{x',x''} )  dV'   dV''\,,\label{equ:totpowerz}
\end{eqnarray}
where
\begin{eqnarray}
\nonumber
I(\mathbf{x',x''} ) &=& \int     e^{ -iw(\mathbf{\hat{x}-\hat k})\cdot (\mathbf{x'}-\mathbf{x''}) } \frac{d\hat{x}d\hat{y}}{\hat{z}} \\
&=&\int_{\hat{x}=-1}^{\hat{x}=1}\int_{\hat{y}=-1}^{\hat{y}=1}e^{ -iw\hat{x}(x'-x'') -iw\hat{y}(y'-y'')} e^{ -iw(\hat{z}-1)(z'-z'')} \frac{d\hat{x}d\hat{y}}{\hat{z}}\,.\label{equ:inte}
\end{eqnarray}
When the wavelength of the photon ($\sim 1/w$) is much smaller than the scale of the source ($L$), i.e. $wL\gg1$, there is a resonance in the region of 
\begin{eqnarray}
wL(1-\hat{z}) = wL(1-\cos\theta_{}) \ll 2\pi \,,
\end{eqnarray}
which means 
\begin{equation}
1-\cos\theta \approx 1-1+\frac{\theta^2}{2}  \ll \frac{2\pi}{wL}
\end{equation}\,,
i.e.
\begin{equation}
\theta \ll \sqrt{\frac{4\pi}{wL}} \,,\quad \text{or} \quad \hat{z} \approx 1 \,,
\end{equation}
which is very similar to the main lobe of the antenna. 
Then the integral in Equ.(\ref{equ:inte}) can be approximated by
\begin{eqnarray}
\nonumber
I(\mathbf{x',x''} ) &\approx& \frac{1}{w^2}\int_{\hat{w}=-\infty}^{\hat{w}=\infty}e^{ -i\hat{w}(x'-x'')}d\hat{w} \int_{\hat{w}=-\infty}^{\hat{w}=\infty} e^{ -i\hat{w}(y'-y'')}  d\hat{w}\\
 &=& \frac{4\pi^2}{w^2}\delta(x'-x'')\delta(y'-y'')\,.\label{equ:inte2}
\end{eqnarray}

The total power from Equ.(\ref{equ:totpowerz}) is
\begin{eqnarray}
\nonumber
P&=& \frac{1}{8} \sum_{i=1}^{3}\int   f_i(x',y',z')f_i^*(x'',y'',z'')  \delta(x'-x'')\delta(y'-y'') dx'dy'dz' dx''dy''dz''\\
\nonumber
&=& \frac{1}{8}\sum_{i=1}^{3}\int   f_i(x',y',z')f_i^*(x',y',z'')
 dx'dy' dz'dz''\\
&=&\frac{1}{8}\sum_{i=1}^{3}\int \bigg[ F_i(x',y')F_i^*(x',y')\bigg]dx'dy' \,,
\end{eqnarray}
where we have defined
\begin{eqnarray}
F(x',y') \equiv \int_{-L}^{L} f_i(x',y',z') dz' \,.  
\end{eqnarray}
In the case of a static electric background, by taking $f_x= 2w \bar E |e_{11}|$, we recover the result with $L=L_z/2$:
\begin{eqnarray}
P = \frac{1}{2}w^2\bar E^2 |e_{11}|^2L_z^2L_xL_y\,.
\end{eqnarray}

For the $h_{11}=-h_{22}$ mode, by using the current components (\ref{equ:zh11jx1}-\ref{equ:zh11jz1}), we get 
\begin{eqnarray}
&&\int_{-L}^{L} f_{x'}(x',y',z')dz' =-2|e_{11}|\int_{-L}^{L} (iw\bar B_{y'} + \partial_{z'} \bar  B_{y'})dz'
%\nonumber
% &=&-2|e_{11}|\bigg( iw\int_{-L}^{L} \bar B_{y'}dz' + \bar  B_{y'}\bigg|_{z'=-L}^{z'=L}\bigg)\\
=-12|e_{11}|B_0r_0^3L  \frac{y'}{(x'^2+y'^2+L^2)^{5/2}} \,,\\
&&\int_{-L/2}^{L/2} f_{y'}(x',y',z')dz' =-12|e_{11}|B_0r_0^3L  \frac{x'}{(x'^2+y'^2+L^2)^{5/2}}\,,\\
&&\int_{-L}^{L} f_{z'}(x',y',z')dz' =  4|e_{11}|\int_{-L}^{L}\partial_{x'} \bar B_{y'} dz'=-60|e_{11}|B_0r_0^3\int_{-L}^{L}\frac{x'y'z'}{|\mathbf{x'}|^7} dz'= 0\,.
\end{eqnarray}
And then
\begin{eqnarray}
\nonumber
P&=& 18|e_{11}|^2B_0^2r_0^6L^2 \int \frac{x'^2+y'^2}{(x'^2+y'^2+L^2)^{5}} dx'dy' \,,\\
\nonumber
&=& 18|e_{11}|^2B_0^2r_0^6L^2 \int_0^R \frac{\rho^2}{(\rho^2+L^2)^5}\rho d\rho \int_0^{2\pi}  d\psi \,, \\
&=& 24\pi |e_{11}|^2B_0^2 L_z^2\left(\frac{r_0}{L_z}\right)^6\left[ 1-\frac{1+4\xi^2}{(1+\xi^2)^4} \right] \,, \label{equ:pz11}
\end{eqnarray}
where we have replaced $L=L_z/2$ and defined 
$\xi = R/L_z$, $R=\sqrt{L_x^2+L_y^2}/2$.

For the $h_{12}=h_{21}$ mode, one can see that Equ.(\ref{equ:zh12jx1}) has the same formalism as (\ref{equ:zh11jy1}) up to a minus sign, and Equ.(\ref{equ:zh12jy1}) is the same as Equ.(\ref{equ:zh11jx1}) since $k_x=k_y=0$. The only difference is the integral of function $f_z$:
\begin{eqnarray}
\nonumber
&&\int_{-L}^{L} f_{z'}(x',y',z')dz' =2|e_{11}| \int_{-L}^{L} ( \partial_{y'}  \bar B_{y'}- \partial_{x'}  \bar B_{x'}) dz'\\
&=&30|e_{11}|B_0 r_0^3 \int_{-L}^{L}  \frac{z}{|\mathbf{x}|^5}\left(\frac{x'^2-y'^2}{|\mathbf{x}|^2}\right) dz' = 0 \,.
\end{eqnarray}
So the total power is the same as that for the $h_{11}$ mode.

\subsubsection{A gravitational wave propagates along the $y$ (or $x$) direction}
In this case, $k_y = w$ and $\hat k_y = 1$. Take the same approach as that in Sec.\ref{sec:gwz} to get the total power radiated approximately
\begin{eqnarray}
P &=&-\frac{1}{8}\sum_{i=1}^{3}\int \bigg[ F_i(x',z')F_i^*(x',z')\bigg]dx'dz' \,,\label{equ:totpowery}
\end{eqnarray}
where
\begin{eqnarray}
F(x',z') \equiv \int_{-L}^{L} f_i(x',y',z') dy' \,.  
\end{eqnarray}
The minus sign in front of Equ.(\ref{equ:totpowery}) coming from the transformation $d\hat y/\hat{z} = - d\hat z/\hat{y}$. 

For the $h_{11}=-h_{33}$ mode, by using the current components (\ref{equ:zh11jx1}-\ref{equ:zh11jz1}), we get
\begin{eqnarray}
\nonumber
&&\int_{-L}^{L} f_{x'}(x',y',z')dy' =2|e_{11}| \int_{-L}^{L} ( iw\bar B_{z'} + \partial_{y'}  \bar B_{z'}) dy'\\
&=&\frac{4iw|e_{11}|B_0r_0^3L}{(x'^2+z'^2)^2(x'^2+z'^2+L^2)^{1/2}}\bigg[\left(3-\frac{L^2}{x'^2+z'^2+L^2}\right)z'^2 - (x'^2+z'^2)\bigg]\,,\\
\nonumber
&&\int_{-L}^{L} f_{y'}(x',y',z')dy'= -4|e_{11}|\int_{-L}^{L} \partial_{x'}\bar B_{z'}  dy'\\
&=&\frac{-8|e_{11}|B_0r_0^3L x'}{(x'^2+z'^2)^{2}(x'^2+z'^2+L^2)^{3/2}}t \bigg[ 3(x'^2+z'^2)+2L^2- z'^2\left(15+\frac{5L^2(x'^2+z'^2)+8L^4}{(x'^2+z'^2)(x'^2+z'^2+L^2)} \right)\bigg]\,,\\
\nonumber
&&\int_{-L}^{L} f_{z'}(x',y',z')dy' =2|e_{11}|\int_{-L}^{L} (iw\bar B_{x'} + \partial_{y'}  \bar B_{x'}  ) dy'\\
&=&\frac{4iw|e_{11}|B_0r_0^3L x'z'}{(x'^2+z'^2)^2}  \frac{3(x'^2+z'^2)+2L^2}{(x'^2+z'^2+L^2)^{3/2}}\,.
\end{eqnarray}
for $x'^2+z'^2 >0$. For the line $x'^2+z'^2 = 0$, we have
\begin{eqnarray}
	&&\int_{-L}^{L} f_{x'}(x',y',z')dy' =2|e_{11}|\bar  B_{z'}\bigg|_{y'=-L}^{y'=L} =-4|e_{11}| B_0 \frac{r_0^3}{L^3}\,,
\end{eqnarray}
and
\begin{eqnarray}
	\int_{-L}^{L} f_{y'}(x',y',z')dy'=\int_{-L}^{L} f_{z'}(x',y',z')dy' = 0\,.
\end{eqnarray}
Thus, the contribution from line $x'^2+z'^2=0$ can be neglected. In the following we assume that the gravtion can travel in the $z-x$ plane with a shortest radius $\epsilon$, i,e, $x'^2+y'^2\geq \epsilon^2$. 

 After transforming $z'=\rho\cos\psi, x'=\rho\sin\psi$, we get $dx'dz' = -\rho d\rho d\theta $. Then it leads to
\begin{eqnarray}
\nonumber
&&\int_\epsilon^R\rho d\rho \int_0^{2\pi} F_x^2 d\psi =\int_\epsilon^R\rho d\rho\frac{16w^2|e_{11}|^2B_0^2r_0^6L^2}{\rho^4(\rho^2+L^2)}\int_0^{2\pi} \bigg[\left(3-\frac{L^2}{\rho^2+L^2}\right)\cos^2\psi - 1\bigg]^2d\psi\\
&&= 4\pi w^2|e_{11}|^2B_0^2r_0^6L^2\int_\epsilon^R \frac{4L^4+12L^2\rho^2+11\rho^4}{\rho^3(\rho^2+L^2)^3} d\rho = -4\pi w^2|e_{11}|^2B_0^2r_0^6L^2\left(\frac{8 L^2+11 \rho ^2}{4 \rho ^2 \left(L^2+\rho ^2\right)^2}\right)\bigg|_ {\rho=\epsilon}^{\rho=R}\,,
\end{eqnarray}

\begin{eqnarray}
\nonumber
&&\int_\epsilon^R\rho d\rho \int_0^{2\pi} F_y^2 d\psi = 64|e_{11}|^2B_0^2r_0^6L^2\int_\epsilon^R \frac{ d\rho}{\rho^5(\rho^2+L^2)^3} \int_0^{2\pi}\sin^2\psi\bigg[ 3\rho^2+2L^2- \rho^2\cos^2\psi\left(15+\frac{5L^2\rho^2+8L^4}{\rho^2(\rho^2+L^2)} \right)\bigg]^2 d\psi \\
\nonumber
&&= 8\pi |e_{11}|^2B_0^2r_0^6L^2\int_\epsilon^R \frac{ 32 L^8+160 L^6 \rho ^2+320 L^4 \rho ^4+300 L^2 \rho ^6+117 \rho ^8 }{\rho^5(\rho^2+L^2)^5}d\rho\\
&&=-8\pi |e_{11}|^2B_0^2r_0^6L^2\left( \frac{64 L^6+256 L^4 \rho ^2+339 L^2 \rho ^4+156 \rho ^6}{8 \rho ^4 \left(L^2+\rho ^2\right)^4} \right)\bigg|_ {\rho=\epsilon}^{\rho=R}
\end{eqnarray}
and
\begin{eqnarray}
\nonumber
&&\int_\epsilon^R\rho d\rho \int_0^{2\pi} F_z^2 d\psi = 16w^2|e_{11}|^2B_0^2r_0^6L^2\int_\epsilon^R\rho d\rho \int_0^{2\pi}  \frac{(3\rho^2+2L^2)^2}{\rho^4(\rho^2+L^2)^{3}}\sin^2\psi\cos^2\psi d\psi \\
\nonumber
&=& 4\pi w^2|e_{11}|^2B_0^2r_0^6L^2\int_\epsilon^R \frac{(3\rho^2+2L^2)^2}{\rho^3(\rho^2+L^2)^{3}} d\rho 
= -4\pi w^2|e_{11}|^2B_0^2r_0^6L^2\left(\frac{8 L^2+9 \rho ^2}{4 \rho ^2 \left(L^2+\rho ^2\right)^2}\right)\bigg|_ {\rho=\epsilon}^{\rho=R}\,.
\end{eqnarray}
Then the total power is
\begin{eqnarray}
\nonumber
P&=& -\frac{\pi}{2} w^2|e_{11}|^2B_0^2r_0^6L^2\left(\frac{4L^2+5 \rho ^2}{ \rho ^2 \left(L^2+\rho ^2\right)^2}\right)\bigg|_ {\rho=\epsilon}^{\rho=R} \\
&-& \pi |e_{11}|^2B_0^2r_0^6L^2\left( \frac{64 L^6+256 L^4 \rho ^2+339 L^2 \rho ^4+156 \rho ^6}{8 \rho ^4 \left(L^2+\rho ^2\right)^4} \right)\bigg|_ {\rho=\epsilon}^{\rho=R}\,,\label{equ:totpowery11}
\end{eqnarray}
where the second term  could be neglect when $w\epsilon \gg 1$. 

For the $h_{13}=h_{31}$ mode, one can see that
 Equ.(\ref{equ:zh13jx2}) has the same formalism as (\ref{equ:zh11jz2}) up to a minus sign, and Equ.(\ref{equ:zh13jz2}) is the same as Equ.(\ref{equ:zh11jx2}). The only difference is the integral of the function $f_y$:
\begin{eqnarray}
\nonumber
&&\int_{-L}^{L} f_{y'}(x',y',z')dy' =2|e_{11}| \int_{-L}^{L} ( \partial_{x'}  \bar B_{x'}- \partial_{z'}  \bar B_{z'}) dy'\\
&=& \frac{4|e_{11}|B_0 r_0^3Lz'}{(x'^2+z'^2)^{2}(x'^2+z'^2+L^2)^{3/2}} \\
\nonumber
&&\cdot \bigg[6(x'^2+z'^2)+4L^2 + \frac{(z'^2-x'^2)(15(x'^2+z'^2)^4+20L^2(x'^2+z'^2)+8L^4)}{(x'^2+z'^2)(x'^2+z'^2+L^2)}\bigg]\,,
\end{eqnarray}
and then
\begin{eqnarray}
\nonumber
&&\int_\epsilon^R\rho d\rho \int_0^{2\pi} F_y^2 d\psi =\int_\epsilon^R d\rho\frac{16|e_{11}|^2B_0^2L^2 r_0^6}{\rho^5(\rho^2+L^2)^{3}}  \int_0^{2\pi}\cos^2\psi\bigg[6\rho^2+4L^2 + \frac{(\cos^2\psi-\sin^2\psi)(15\rho^4+20L^2\rho^2+8L^4)}{(\rho^2+L^2)}\bigg]^2 d\psi\\
\nonumber
&=&8\pi|e_{11}|^2B_0^2L^2 r_0^6\int_\epsilon^R \frac{\left(160 L^8+800 L^6 \rho ^2+1552 L^4 \rho ^4+1380 L^2 \rho ^6+477 \rho ^8\right)}{\rho^5(\rho^2+L^2)^{5}}d\rho\\
\nonumber
&=&-8\pi|e_{11}|^2B_0^2L^2 r_0^6\bigg(\frac{320 L^{10}+1280 L^8 \rho ^2+1939 L^6 \rho ^4+1468 L^4 \rho ^6+672 L^2 \rho ^8+192 \rho ^{10}}{8 L^4 \rho ^4 \left(L^2+\rho ^2\right)^4}
+\frac{24}{L^6}\log\frac{\rho^2+L^2}{\rho^2}\bigg)\bigg|_ {\rho=\epsilon}^{\rho=R}\,.
\end{eqnarray}
Therefore, the total power is the same as that with $h_{11}$ mode since the above contribution can be neglected when $w\epsilon\gg1$.

\section{Detail calculations in the near-zone of  a neutron star} \label{sec:app2}
To calculate Equ.(\ref{equ:iomeg}), one needs to perform the following integration
\begin{eqnarray}
\int e^{-iw\theta(r'\theta'-r''\theta'') } e^{-iw\sin\theta\phi(\sin\theta'\phi' r'-\sin\theta'' \phi'' r'') } dr'dr'' r'^2r''^2\sin\theta' d\theta' d\phi'\sin\theta'' d\theta'' d\phi'' d\Omega\,.
\end{eqnarray}
By using the short wave approximation and the following transformations
\begin{eqnarray}
x = r\theta\,,\quad y =r\sin\theta \phi \,,\quad 
dx \approx rd\theta \,,\quad  dy \approx r\sin\theta d\phi  \,,
\end{eqnarray}
and
\begin{eqnarray}
\hat x = \theta\,,\quad \hat y =\sin\theta \phi \,,\quad 
d\hat x \approx d\theta \,,\quad  d\hat y \approx \sin\theta d\phi  \,,
\end{eqnarray}
we get
\begin{eqnarray}
	\nonumber
&&\int d\hat xd\hat y e^{-iw \hat x(x'-x'')} e^{-iw\hat y(y'-y'')} dx'dy' dx''dy''  \\
&\approx&
\frac{4\pi^2}{w^2} \int \delta(x'-x'')\delta(y'-y'') dx'dy' dx''dy''
= \frac{4\pi^2}{w^2} \int dx'dy' = \frac{4\pi^2}{w^2} \int r'^2\sin \theta'  d\theta' d\phi' \,.
\end{eqnarray}
To calculate the radial part in the integrals Equs.(\ref{equ:iomeg1}) and (\ref{equ:iomeg2}), the integral variable is changed to $\eta = 2M/r$ in the following: 
\begin{eqnarray*}
	&&\int  r'^4N^{-2} K_1G^*(r') dr' = \frac{3}{4}  B_0R^4\cos\chi  \int   \frac{|K_1|^2}{M^2r'}\bigg( \frac{r'}{M}\ln N^2+\frac{1}{N^2}+1\bigg)  dr'\\
	&=&\frac{3}{4M^3}  B_0R^4\cos\chi  \int   \left(w^2+\frac{1}{r'^2}\right) \bigg[ \ln N^2+\frac{M}{r'}\left(\frac{1}{N^2}+1\right) \bigg]  dr'\\
	&=&\frac{3}{4M^3}  B_0R^4\cos\chi  \int   \left(w^2+\frac{1}{4M^2}\eta^2\right) \bigg[ \ln (1-\eta)+\frac{1}{2}\eta \left(\frac{1}{1-\eta}+1\right) \bigg]\frac{-2M}{\eta^2}  d\eta\\
	&=& \frac{3}{4M^4}  B_0R^4\cos\chi 
	\left[ \frac{3M}{2r}
	-\frac{M^2}{2 r^2}+ \left(\frac{3}{4}-w^2 M^2
	+\frac{r M^2w^2 }{M} -\frac{M}{ r} \right)\ln \left(1-\frac{2 M}{r}\right)
	\right]\bigg|_R^{R+L}\\
	&\approx& \frac{3}{4M^4}  B_0R^4\cos\chi \bigg[ -2 -\frac{1}{6} \frac{4M^2}{r^2}-\frac{1}{6} \frac{8M^3}{r^3} \bigg] \left(M^2 w^2\right)\bigg|_R^{R+L}\\
	&\approx& \frac{w^2}{8M^2}  B_0R^4\cos\chi \bigg[ \frac{4M^2}{R^2}(1-(1+L/R)^{-2})+ \frac{8M^3}{R^3}(1-(1+L/R)^{-3}) \bigg]\\
	&=&  w^2 \cos\chi B_0L R \left(1 + \frac{3M}{R}\right)\,,
\end{eqnarray*}
and
\begin{eqnarray*}
	&&\int  N^{-1} K_0^*G(r'') dr'' = 
	\frac{3}{4}  B_0R^4\cos\chi  \int   \frac{|K_0|^2}{M^2r''^5}\bigg( \frac{r''}{M}\ln N^2+\frac{1}{N^2}+1\bigg)  dr''\\
	&=&\frac{3}{4M^3}  B_0R^4\cos\chi  \int   \frac{w^2}{r''^4} \bigg[ \ln N^2+\frac{M}{r''}\left(\frac{1}{N^2}+1\right) \bigg]  dr''\\
	&=& \frac{3}{4M^3}  B_0R^4\cos\chi  \int \frac{w^2}{(2M)^4}\eta^4 \bigg[ \ln (1-\eta)+\frac{1}{2}\eta \left(\frac{1}{1-\eta}+1\right) \bigg]\frac{-2M}{\eta^2}  d\eta \\
	&=& -\frac{3w^2}{4M^3} \frac{1}{(2M)^3} B_0R^4\cos\chi  \int \eta^2\bigg[ \ln (1-\eta)+\frac{1}{2}\eta \left(\frac{1}{1-\eta}+1\right) \bigg] d\eta \\
	&=& \frac{3w^2B_0R^4\cos\chi }{(2M)^6}  \left[ \frac{10M}{3r}+\frac{5 (2M)^2}{6r^2}+\frac{5 (2M)^3}{9r^3}-\frac{(2M)^4}{4r^4}+ \left(\frac{5}{3}-\frac{2(2M)^3}{3r^3}  \right)  \ln (1-\frac{2M}{r})\right]\bigg|_R^{R+L} \\
	&\approx& \frac{3w^2B_0R^4\cos\chi }{(2M)^6}  \left(\frac{(2M)^6}{18r^6}+\frac{(2M)^7}{14r^7}\right)\bigg|_R^{R+L} \\
	&\approx& 3w^2B_0R^4\cos\chi   \left(\frac{1}{18R^6}(1-(1+L/R)^{-6})+\frac{M}{7R^7}(1-(1+L/R)^{-7})\right)\\
	&\approx& \frac{w^2B_0L\cos\chi }{R^3}  \left(1+\frac{3M}{R}\right)\,,
\end{eqnarray*}
where we only keep the leading order of $M/R$, and take the approximation $L\ll R$ in the final results of the above integrals.

\bibliographystyle{unsrt}
\bibliography{MagneticFieldofPulsar}

\end{document}